\documentclass{IEEEcsmag}

\usepackage[colorlinks,urlcolor=blue,linkcolor=blue,citecolor=blue]{hyperref}
\expandafter\def\expandafter\UrlBreaks\expandafter{\UrlBreaks\do\/\do\*\do\-\do\~\do\'\do\"\do\-}

\usepackage[table]{xcolor} 
\usepackage{upmath,color}
\usepackage{booktabs} 
\usepackage{multicol}
\usepackage{multirow}
\newcommand{\commentout}[1]{}

\jvol{XX}
\jnum{XX}
\paper{8}
\jmonth{Month}
\jname{Publication Name}
\jtitle{Publication Title}
\pubyear{2021}

\newenvironment{myitemize}
    {\begin{itemize}[\setlength{\IEEElabelindent}{\dimexpr-\labelwidth+5pt}
    \setlength{\itemsep}{4pt}
    \setlength{\listparindent}{\parindent}
    ]}{\end{itemize}}

\begin{document}

\sptitle{Defining A New Cross Reality: Digital Twins and Mixed Reality Worlds}

\title{Cross-Reality Lifestyle: Integrating Physical and Virtual Lives through Multi-Platform Metaverse}

\author{Yuichi Hiroi}
\affil{Cluster Metaverse Lab, Tokyo, 141-0031, Japan}

\author{Yuji Hatada}
\affil{The University of Tokyo, Tokyo, 113-8656, Japan}

\author{Takefumi Hiraki}
\affil{Cluster Metaverse Lab, Tokyo, 141-0031, Japan and University of Tsukuba, Ibaraki, 305-8550, Japan}

\markboth{Defining A New Cross Reality: Digital Twins and Mixed Reality Worlds}{Defining A New Cross Reality: Digital Twins and Mixed Reality Worlds}

\begin{abstract}\looseness-1Technological advances are redefining the relationship between physical and virtual space. Traditionally, when users engage in virtual reality (VR), they are completely cut off from the physical space. Similarly, they are unable to access virtual experiences while engaged in physical activities. However, modern multi-platform metaverse environments allow simultaneous participation through mobile devices, creating new opportunities for integrated experiences. This study introduces the concept of "cross-reality lifestyles" to examine how users actively combine their physical and virtual activities. We identify three patterns of integration: 1) Amplification: one space enhances experiences in the other; 2) Complementary: spaces offer different but equally valuable alternatives; and 3) Emergence: simultaneous engagement creates entirely new experiences. We propose the ACE Cube framework that analyzes these patterns as continuous characteristics, and by integrating this analysis with technical requirements of commercial platforms, we provide practical guidelines for platform selection, technical investment prioritization, and cross-reality application development.
\end{abstract}

\maketitle

\chapteri{T}he relationship between physical and virtual spaces is being redefined by advances in information technology. Since the era of video games and early social virtual worlds like Second Life, these spaces have been treated as mutually exclusive: users in virtual spaces were isolated from the physical information, while virtual experiences remained disconnected from everyday life. Recent developments in immersive VR technology have intensified this separation by creating a stronger presence and an embodied avatar experience~\cite{Slater2022-mr}. VR creates a sense of being in a completely different place, enabling the simulation of extraordinary, non-everyday experiences. This separation shaped the technological development in two different directions. Digital twin technology~\cite{Botin-Sanabria2022-zd} positions virtual spaces as tools for simulating and optimizing real environments, establishing physical space as primary. In contrast, metaverse platforms~\cite{Ritterbusch2023-ct} such as VRChat\footnote{\url{https://hello.vrchat.com/}} and Cluster\footnote{\url{https://cluster.mu/en}} prioritize virtual worlds independent of physical constraints, creating new venues for unconstrained self-expression~\cite{Freeman2021-ad}.

\begin{figure}[t]
    \centering
    \includegraphics[width=1\linewidth]{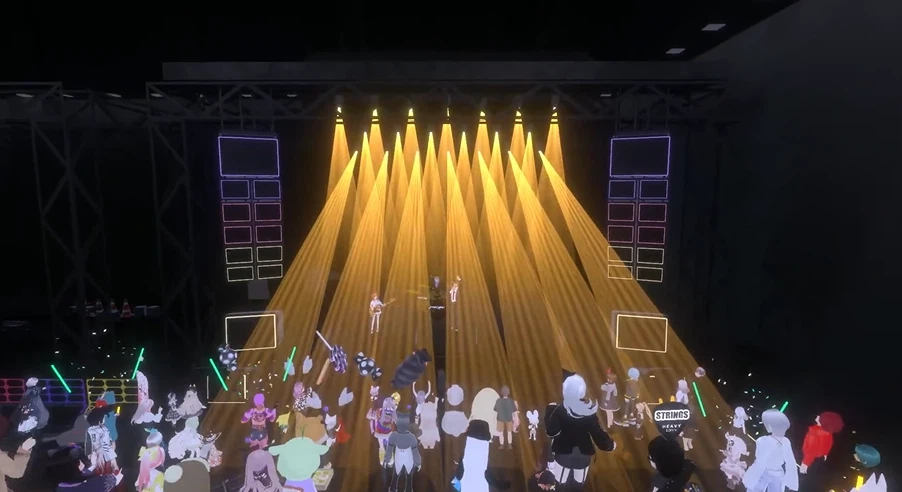}
    \caption{A live music event on Cluster, a metaverse platform accessible from smartphones. Although some users are on their smartphones, they are dancing and responding to the music in VR space.}
    \label{fig:cluster-music-jam}
\end{figure}

However, current trends suggest a shift beyond this dualistic division. For example, younger generations routinely blend physical activities with virtual social connections, such as doing chores while staying connected through video calls~\cite{Suh2018-ai}. The coexistence of activities in real and virtual spaces has expanded through metaverse development, especially as platforms adopt cross-device accessibility from mobile devices, such as VR-HMD and smartphones, and IoT device integration. For example, Roblox\footnote{\url{https://www.roblox.com/}}, one of the world's largest non-immersive metaverse platforms added VR-HMD support in 2023, while VRChat, which is played by many VR-HMD users, introduced smartphone access in 2024. As diverse access routes to these spaces have been developed, users simultaneously engage in both spaces, participating in metaverse events while performing physical activities (Fig.~\ref{fig:cluster-music-jam})\footnote{Cluster Music Jam Vol.1. \url{https://cluster.mu/e/96cf9bff-5cd5-406b-ac50-dc51e9ef08a0}}. 
Unlike previous studies that have focused mainly on the master-slave relationship that exists between the two spaces, such as real-world VR training~\cite{Slater2016-tp} or user activity on metaverse platforms~\cite{Freeman2021-ad}, these emerging patterns of simultaneous engagement remain understudied.

In this paper, we introduce the concept of “cross-reality life experiences” to describe emerging lifestyle patterns from seamless physical-metaverse fusion. We define these experiences as novel activities where neither space dominates and life in both realms integrates into unified activity. 
To overcome traditional virtual-real opposition, we reinterpret these as interaction patterns between multiple "verses"—different implementations of the same activity space.



We analyze these cross-reality life experiences as integration patterns between different types of verses (e.g., physical and metaverse) through three dimensions: Amplification, Complementarity, and Emergence.
Rather than considering them as discrete categories, we conceptualize these dimensions as continuous characteristics, each of which is exhibited by experiences that vary in intensity across all dimensions. We map these experiences within a three-dimensional ACE cube and position commercial metaverse platforms alongside their technical requirements. This establishes connections between the experiences' integration patterns and technical implementation. This multidimensional framework transcends the physical-virtual binary opposition and provides practical guidelines for selecting platforms, prioritizing technical investments, and developing cross-reality applications.

\begin{figure*}[t]
    \centering
    \includegraphics[width=1\linewidth]{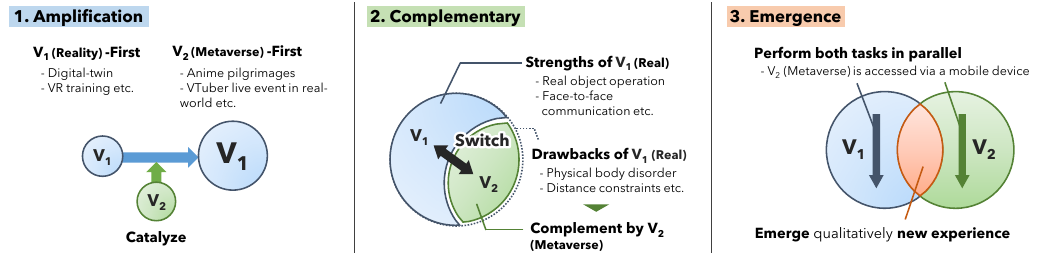}
    \caption{Patterns of integration between verse 1 (V$_{1}$, physical reality) and verse 2 (V$_{2}$, metaverse) spaces. The diagram illustrates three distinct patterns: (1) Amplification, where one space catalyzes value enhancement in the other through both reality-first and metaverse-first approaches; (2) Complementary, where the one verce provides alternative options to overcome anoter verse's inherent limitations while maintaining another verse's core strengths; and (3) Emergence, where parallel engagement in both spaces creates qualitatively new experiences beyond simple combination.}
    \label{fig:lifestyle-type}
\end{figure*}

\section{UNIFIED ANALYSIS THROUGH ACE CUBE}
Milgram et al.'s Reality-Virtuality Continuum~\cite{Milgram1995-ku} positioned physical and virtual spaces as opposing poles on a one-dimensional spectrum, with Augmented Reality (AR) augmenting physical space with virtual information, and Augmented Virtuality (AV) augmenting virtual space with physical elements filling the intermediate space. While this framework focused primarily on perceptual experiences, our research extends it to everyday activities.

Zeltzer's AIP cube~\cite{Zeltzer1992-ku} pioneered three-dimensional VR system analysis. The AIP cube is a three-dimensional framework that evaluates technical characteristics of VR systems along three axes: Autonomy, Interaction, and Presence axes, with continuous values from 0 to 1. The coordinate (1, 1, 1) represents the "ultimate VR system".

\begin{figure}[t]
    \centering
    \includegraphics[width=1\linewidth]{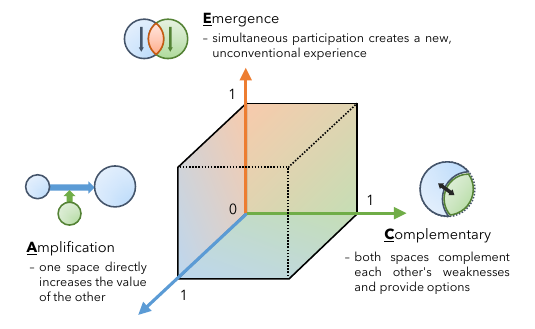}
    \caption{The ACE Cube Framework is used to analyze cross-reality experiences. This three-dimensional framework maps cross-reality experiences along the Amplification (A), Complementarity (C), and Emergence (E) axes.}
    \label{fig:ace-cube}
\end{figure}

We propose a new three-dimensional framework called the "ACE cube" that analyzes value creation in users' daily-life experiences, while the AIP cube measures technical system capabilities. We analyze integration patterns between physical and metaverse spaces as continuous characteristics along three dimensions (Fig.~\ref{fig:lifestyle-type}):
\begin{myitemize}
    \item \textbf{A-axis (Amplification)} measures how experiences in one space enhance capabilities in another. For example, VR training improves real surgical skills, and real-life events deepen fan activities for virtual YouTubers (VTubers). From a temporal perspective, amplification-type integration involves a clear sequence in which activities in one space enhance activities in another.
    \item \textbf{C-axis (Complementary)} quantifies how one space complements the constraints and limitations of the other. This restores experiential wholeness through metaverse-physical fusion, which was previously limited by single-space constraints due to the absence of metaverse options. From a temporal perspective, complementary-type integration is characterized by "simultaneously selectable" states. Multiple spaces exist as equal alternatives, enabling the selection of optimal combinations according to the situation.
    \item \textbf{E-axis (Emergence)} measures qualitatively new values that arise from simultaneous multi-space experiences that exceed simple summation. From a temporal perspective, emergent-type fusion features "persistent simultaneity." While complementary integration recovers wholeness through switching spaces, emergent integration creates new experiences through parallel engagement.
\end{myitemize}


Each cross-reality experience exhibits specific intensities along these axes, represented as continuous coordinates within the ACE cube (Fig.~\ref{fig:ace-cube}). We map experience cases and position existing platforms within this cube. The coordinate (1, 1, 1) represents the "ultimate metaverse platform" that supports all types of experiences.


\subsection{Spatial Relationship Reconstruction Through the Verse Concept}
Traditional analyses have assumed qualitative differences between "real" and "virtual" spaces. However, recent developments in the metaverse reveal that, although technical requirements differ, experiences across spaces are increasingly similar.

To overcome the "virtual-real" opposition, we reinterpret it as an interaction pattern between multiple "verses." Physical and metaverse spaces are different implementations of the same "verse," or activity space. For example, considering hosting a VTuber event,  the physical implementation requires securing a venue, setting up a stage, and installing lighting and sound. In contrast, the metaverse implementation requires designing a virtual venue and constructing a 3D stage with virtual lighting and sound. Although the technologies differ, the essential structure of constructing event space remains identical.

From this perspective, there are multiple "verses" with different implementation technologies that exist in parallel. Each verse has specific constraints and affordances: physical verses face gravity and physical laws but provide tactile feedback and direct embodiment. In contrast, metaverses transcend physical constraints, offering expressive freedom and global accessibility.

The "verse" concept provides a neutral way to describe traditional relationships as "interverse interaction patterns." For example, amplification experiences essentially become uniform value enhancements through interverse interactions. Therefore, we represent both physical and metaverse activities as the same "interverse amplification" phenomena, whether physical activities improve metaverse capabilities or vice versa.
Furthermore, this concept supports future multi-metaverse integration analysis. For example, when avatar creation skills in VRChat enhance event hosting capabilities in Cluster, this can be analyzed as "Metaverse V$_{1} \rightarrow$ V$_{2}$" amplification.





\section{CLASSIFICATION OF CROSS-REALITY LIFE EXPERIENCES}
\begin{figure*}[t]
    \centering
    \includegraphics[width=1\linewidth]{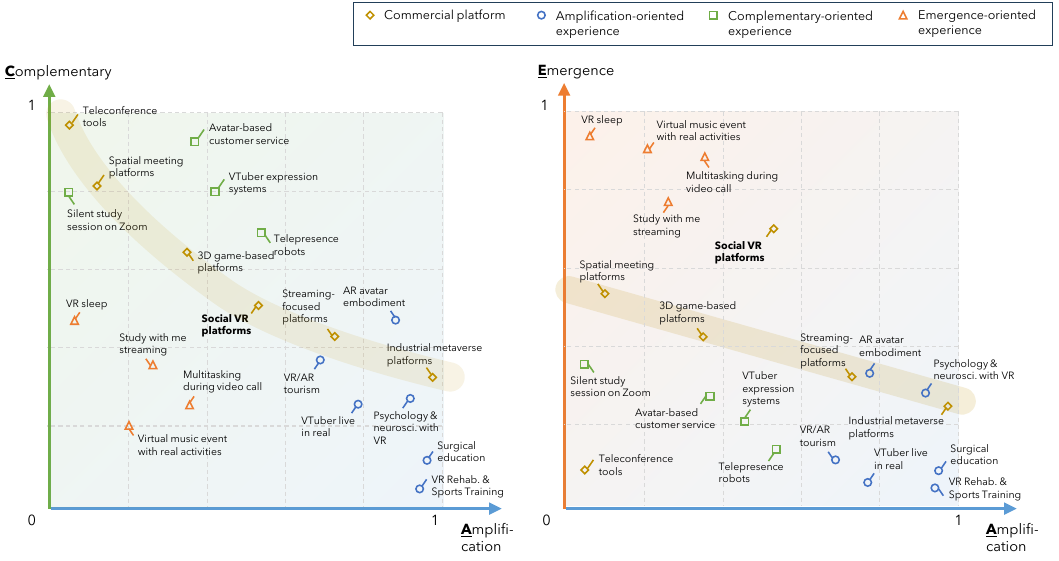}
    \caption{Cross-reality experiences and platform distribution in the ACE Cube. Two-dimensional projections show the A-C (left) and A-E (right) planes. Examples of experiences and commercial platforms are mapped onto these planes. When mapping the platforms, a yellow curve appeared on the A-C plane, indicating that existing platforms have adopted a clear strategic differentiation aligned with the A-C trade-off. However, social VR platforms demonstrate superior emergent capabilities on the A-E plane compared to platforms in other categories.}
    \label{fig:ace-map-experience}
\end{figure*}

Figure~\ref{fig:ace-map-experience} shows examples of cross-reality experiences mapped onto the ACE Cube. Three researchers created this ACE map by surveying over 200 official event cases on the Cluster platform and related work. Currently, this analytical method relies on the judgment of researchers. However, developing more objective mapping methods is a topic for future research.
Based on this map, the following sections examine each type with relevant literature.


\subsection{Amplification-oriented Experiences}
Amplification-type integration involves enhancing the experiential value of one verse through the intervention of the other.

\subsubsection{Reality-First Amplification}
Reality-first amplification builds on experiences in real space, where metaverse elements enhance the inherent value. This pattern is characterized by how partial participation in the metaverse serves as a catalyst to qualitatively enhance real-world experiences. This reality-first amplification has traditionally been considered in the context of digital twins and VR training, and Slater and Sanchez-Vives provide a comprehensive overview~\cite{Slater2016-tp}.

This pattern has shown significant effects across multiple domains. Medical training, particularly surgical education, uses virtual spaces to develop complex skills in reality while ensuring safety~\cite{Abich2021-ht, Jensen2018-dv, Xie2021-cb}. In contrast, functions that complement real surgical techniques are limited, and training and practice do not occur simultaneously. Thus, we plot them on high A and low C and E values.

VR/AR tourism applications combine travel experiences in a virtual environment with real tourism~\cite{Yung2019-km}. While these applications primarily serve amplification roles through AR-enhanced physical experiences and VR-improved travel motivation, they also serve complementary roles in overcoming the distance, cost, and time constraints of the physical world.

In psychology and neuroscience fields, VR research on body ownership has advanced, revealing flexibility in self-recognition that was difficult to observe with traditional methods~\cite{Loomis1999-io, Maister2015-rg, Blanke2015-tb}. Metaverse experiments deepen our understanding of the physical world. For example, researchers have reported that embodying avatars of different races can reduce prejudice and develop empathy in the physical world~\cite{Herrera2018-mj, Bertrand2018-kw}.

AR avatar~\cite{Genay2022-pm}, which overlays virtual avatars on physical bodies through AR displays, brings the Proteus effect~\cite{nick2007yee}, the observation that avatar characteristics influence user behavior and self-perception, into the physical verse.
The Proteus effect allows for the direct application of improved self-perception and behavioral patterns from the metaverse to the physical world. It also enables self-expression embodiment that would otherwise be impossible in physical environments through AR technology.

\subsubsection{Metaverse-First Amplification}
Metaverse-first amplification involves strengthening the values inherent in metaverse by adding physical elements. The most basic example is anime pilgrimage, a cultural phenomenon that involves visits to real locations featured in anime and video games~\cite{Okamoto2015-vw}, which more strongly amplify the reality of metaverse experiences. In addition, offline "meet-ups" that deepen bonds formed in online game communities in the real world exemplify the metaverse space relationships being strengthened through physical interaction~\cite{Sessions2010-qu}. 

Contemporary developments of this amplification pattern include "oshi-katsu" (fan activities) supporting VTubers and streamers~\cite{Liu2023-px}. These activities have evolved into real-world behaviors such as participating in physical events and purchasing merchandise, which amplifies the fan experience.

\subsection{Complementary-oriented Experiences}
Complementary integration restores experiential wholeness by combining physical and metaverse options. Telepresence robots in customer service exemplify this integration by reducing spatial and temporal constraints~\cite{Song2022-hb}. Virtual platforms, such as teleconferencing tools, complement face-to-face interactions by providing alternative modes of engagement.


Avatars enable forms of self-expression that are impossible in physical reality.  VTubers establish a new form of expression that goes beyond the appeal of reality by combining the inner self of a real person with a new attribute expressed through an avatar~\cite{Lu2021-pr} Similarly, CGI-based virtual influencers are emerging as alternatives to traditional human influencers in marketing~\cite{Byun2023-db, Laszkiewicz2023-ws}.

Hatada et al. introduced an avatar system in a cafe where individuals with disabilities provide customer service using telepresence robots~\cite{hatada2024robotcafe}. The robots’ standardized appearance made it difficult for employees to express their individuality, so the system incorporated customizable virtual avatars as an additional option for customer interactions. This  enabled employees to seamlessly switch between physical telepresence robots and virtual avatars, allowing them to provide service from both the real and virtual spaces. The system empowered employees to engage more dynamically and authentically, creating a richer and more fulfilling customer service experience.

Alternative means of face-to-face interaction through virtual environments create usable options simultaneously for real and virtual, enabling optimal combinations according to the situation. "Silent study sessions" on Zoom provide ambient presence without social pressure~\cite{Cho2023-bm}, while screen sharing capabilities can enhance productivity beyond traditional meetings. 



\subsection{Emergence-oriented Experiences}
Emergent integration occurs when activities in the multiple (e.g., real and virtual) verses take place simultaneously and interact. Two technological advances enable this integration: the maturation of social VR communities and the widespread availability of hands-free VR devices, including VR-HMDs and smartphones. These developments make simultaneous engagement in both spaces technically feasible for general users.

The HCI and CSCW community has identified "incomplete" social presence as a significant pattern in which users modulate their level of digital engagement. Research shows that teens actively value this partial social presence, such as multitasking while participating in video calls~\cite{Suh2018-ai}.
"Study with me" videos on YouTube Live demonstrate how users create personalized digital social environments that balance engagement with controllable peer pressure~\cite{Lee2021-ap}.


By combining "incomplete" social presence with non-immersive metaverse experiences via mobile devices, the "multitasking metaverse" experience emerges. As shown in the introduction section, users achieve optimal engagement by placing avatars in virtual spaces while maintaining physical activities. For example, in a virtual music event in the metaverse platform, users study and listen to the music while their avatar automatically dances and actively reacts at key points in the event (Fig.\ref{fig:cluster-music-jam}). This approach satisfies users who want flexible participation without constant full immersion.

Maloney and Freeman's report of “VR sleep”---a phenomenon where individuals sleep alone physically while maintaining a sense of social presence in a virtual environment---illustrates how the fusion of personal physical activities with virtual social interaction creates a unique form of intimacy~\cite{Maloney2020-es}. They note that users often feel as though they are sleeping alongside friends in the virtual world, with moments blurring the boundaries between physical and virtual spaces, resulting in a hybrid experience.


The ultimate research on such emergent experiences includes multiverse body control~\cite{Miura2022-yx}. This involves controlling bodies in multiple metaverses while maintaining a presence in the physical world. Having a single consciousness exist in multiple verses creates revolutionary experiences and suggests the potential transformation of traditional concepts of embodiment and spatiality.

\section{ANALYSIS OF COMMERCIAL PLATFORMS}
We classify current metaverse platforms into six categories based on core functions and technical features, examining how each supports different forms of cross-reality integration.
Note that while platforms may share features across multiple categories, we classify them according to their primary use cases and defining characteristics.

\subsubsection{1. Industrial Metaverse Platforms (e.g. Task-specific Applications, NVIDIA Omniverse)}
These platforms focus on workplace simulation and digital twin creation for the manufacturing and construction industries. 
They integrate CAD data and real-time sensor information to optimize plant layouts and verify operational processes. Technical requirements include precise compatibility with industrial data, secure high-performance infrastructure, and precise control environments through AR/VR devices. Content priorities include accurate physical simulation with real-time capabilities, while the platform must provide enterprise-class reliability and scalability.
These platforms primarily support reality-first amplification through sensor and actuator integration, and complementary integration through remote robot operation. 

\subsubsection{2. Streaming-focused Platforms (e.g., YouTube Live, Twitch)}
These platforms emphasize video streaming and community building around content creators. They require high-quality video delivery and viewer engagement features (chat, donations) supported by a robust infrastructure for large-scale, simultaneous broadcasting. Technical considerations include low-latency, high-quality video delivery, real-time interaction processing, and large-scale viewer data management. Device requirements remain moderate, supporting standard PC and mobile access.
These platforms effectively amplify metaverse content to external audiences. Although some streamer broadcasts contents related to emergence integration, e.g. "study with me" streaming, the primary interaction pattern remains one-way from streamer to viewer. 

\subsubsection{3. 2D Teleconference Tools (e.g., Zoom, Teams, Meet)}
These communication platforms serve the needs of business meetings and online education. They prioritize stable audio and video transmission while offering document sharing and chat. Technical requirements focus on low latency, high security, intuitive user interfaces, and compatibility with standard webcams and microphones. Enterprise applications place a high priority on security and stability. These platforms represent the most mature sector of complementary integration, having established themselves as alternatives to physical meetings and becoming essential for hybrid work and online education.

\subsubsection{4. 2D Spatial Teleconference Platforms (e.g., Gather, SpatialChat)}
These platforms feature avatar movement on 2D maps with proximity-based audio functionality, enabling casual spatial interactions such as poster sessions at academic conferences and social gatherings.
Technical features include 2D avatar control, distance-based audio adjustment, and lightweight browser-based implementation. By minimizing hardware requirements, these platforms enable broad user participation.
As examples of complementary integration, they enhance traditional meeting systems with game-like and avatar elements to promote natural interaction. Continuous use, such as during parallel tasks, can lead to quasi-emergent integrations.

\subsubsection{5. 3D game-based platforms (e.g. Fortnite, Roblox, Minecraft)}
While these systems were originally developed as gaming platforms with predefined rules, they have evolved toward metaverse functionality through enhanced user-generated content (UGC) capabilities and large-scale events. Technical requirements include advanced 3D graphics processing, support for a high number of concurrent connections, and optimization across multiple devices, such as PCs, gaming consoles, and mobile devices. Many platforms offer their own development environments and marketplaces for UGC.

From a cross-reality integration perspective, these platforms specialize in complementary functions. For example, Minecraft and Roblox provide opportunities for large-scale construction and creation of structures that transcend physical laws and would be impossible in reality. These opportunities are difficult to achieve through traditional classroom instruction and serve as entry points for programming education. Additionally, the creative skills gained through gaming partially transfer to real-world capabilities; however, this remains moderate due to gaming's primary focus on gaming experiences. This focus of the platform design also limits the creation of emergent experiences that occur simultaneously with physical activities.

\subsubsection{6. Social VR Platforms (e.g. VRChat, Cluster, SecondLife, VirtualCast, REALITY, ZEPETO)}
These platforms focus on communication and world editing through 3D avatars, supporting diverse applications such as music performances, academic conferences, and community events. Technical requirements include high-quality 3D avatar rendering, multi-platform support (VR/PC/mobile), and advanced body tracking technologies such as motion capture. They offer significant UGC freedom and encourage active user creation of avatars and worlds.
Compared to other platforms, these platforms demonstrate superior capabilities in creating emergent experiences. These platforms host many high-E, or "while participating," experiences, including VR sleep phenomena and virtual music events combined with physical activities.

\begin{figure*}[t]
    \centering
    \includegraphics[width=1\linewidth]{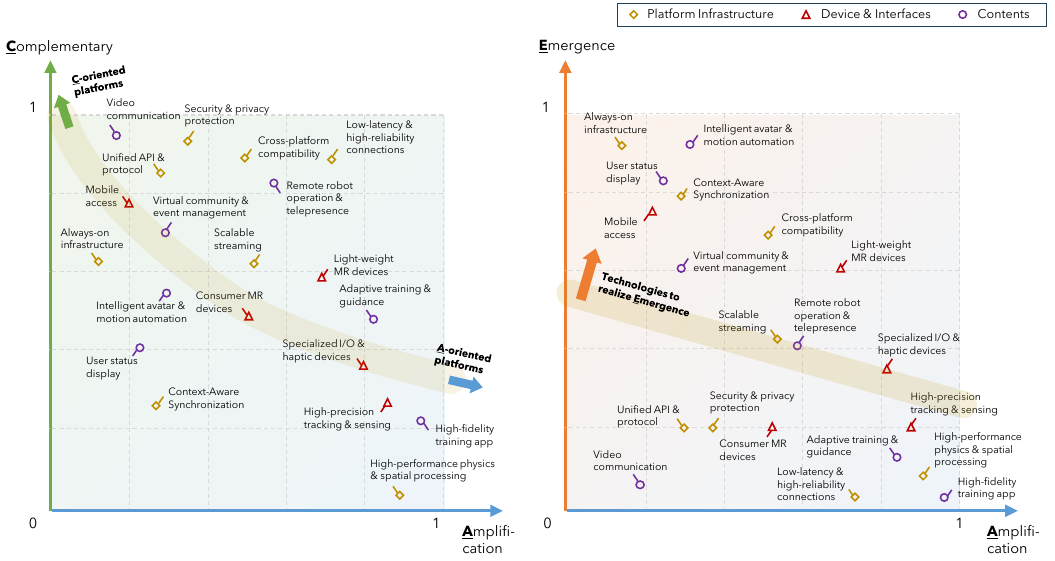}
    \caption{Technical requirements mapping in the ACE cube. Technical requirements are classified as content (circles), platform infrastructure (diamonds), or devices and interfaces (triangles) across projections A-C and A-E. The yellow lines in both planes are in the same position as in Fig.~\ref{fig:ace-map-experience} They represent the technical requirements for realizing cross-reality experiences, as well as platform selection and mapping. In the A-C plane, the yellow line help select platform technology elements according to experience A-C trade-offs. In the A-E plane, the yellow line show that technologies positioned along the E-axis provide emergent experiences through supporting platforms.}
    \label{fig:ace-map-technology}
\end{figure*}

\section{TECHNICAL FRAMEWORK FOR CROSS-REALITY LIFESTYLE}
To clarify the relationship between experience cases and technical requirements, we analyze the technical requirements for a cross-reality experience in the ACE cube. 

These technical requirements are classified into three layers:
\begin{myitemize}
    \item \textbf{Contents:} Experience design methodology and interaction in metaverse spaces, including UGC creation and sharing systems, integration of immersive social events, and flexible participation models that adapt to user engagement levels.
    \item \textbf{Platform Infrastructure:} Core technical infrastructure including multi-device support, real-time communication systems, scalability for multiple users, unified 2D / 3D space management and persistent data storage.
    \item \textbf{Device \& Interface:} Technologies bridging physical and virtual spaces through sensing, motion capture, VR / AR devices and physical feedback systems.
\end{myitemize}

\subsection{Technical Requirements Mapping in ACE Cube}
Figure~\ref{fig:ace-map-technology} shows the ACE coordinate values for each technical requirement. This analysis clarifies the technical conditions necessary to realize specific experience domains. 

\begin{myitemize}
\item \textbf{High-A technologies} (e.g., high-precision physics and spatial processing, high-precision tracking and sensing, and cross-platform compatibility) significantly enhance understanding and predictive capabilities through precise reality reproduction, supporting capability enhancement experiences such as medical training and VR rehabilitation. 
\item \textbf{High-C technologies} (e.g., video communication, unified APIs and protocols, and security and privacy protection) serve as direct substitutes for physical face-to-face interaction by resolving distance, time, and technical compatibility constraints and enabling complementary experiences such as teleconferencing. 
\item \textbf{High-E technologies} (e.g., avatar and motion automation, always-on infrastructure, mobile access, and context-aware synchronization systems) automate virtual presence during physical activities, enabling "parallel participation" and providing the basis for emergent experiences.
\end{myitemize}

\subsection{Platform Design Guidelines through Technology Integration}
A comparison of the ACE analysis of existing platforms and the mapping of technical requirements reveals the factors that differentiate platforms from each other.

Social VR platforms significantly surpass others in their ability to support emergent experiences due to their integrated implementation of advanced technologies, such as avatar and motion automation, always-on infrastructure, and context-aware synchronization systems. These technologies provide the foundation for users to create new and unexpected experiences.

Industrial metaverse platforms specialize in high-precision simulation by focusing on advanced technologies, such as high-precision physics, spatial processing, and specialized input/output devices. These technologies enable experiences that significantly improve real-world operational efficiency.
Teleconferencing tools focus on video communication technologies and security and privacy protection to serve as direct substitutes for physical constraints.

Emerging technologies, such as lightweight mixed reality (MR) devices, enable improved sustained use capabilities by reducing the burden of wearing them and by realizing new lifestyles that integrate virtual elements into daily life through extended, simultaneous experiences. These technologies are expected to pioneer new experience domains by integrating traditionally exclusive high-ACE regions.

This analysis helps developers identify the technical requirements for the target experience domains and formulate strategies to differentiate their platforms from existing ones. Developing platforms that approach the level of ACE Cube (1, 1, 1) requires integrated development in all technology domains.


\section{CONCLUSION}
This paper introduced cross-reality lifestyles and the ACE Cube framework analyzing physical-virtual space integration. We introduced the "verse" concept, treating both spaces as activity spaces with different implementations, and analyzed integration patterns as continuous characteristics across Amplification, Complementary, and Emergence dimensions.
The case analysis mapped experiences and technical requirements within the ACE space, establishing connections between integration patterns and implementation. Platform analysis revealed strategic differentiation along the A-C axis, with social VR platforms showing superior emergent capabilities.
The ACE Cube framework provides guidelines for selecting platforms, prioritizing technical investments, and developing applications. Developers can identify technical requirements for target domains and formulate strategies for the ultimate metaverse platform.

Future research should focus on developing objective evaluation methods, exploring new interaction paradigms, and integrating high-ACE domains through emerging technologies. We expect this framework to guide the technical design of future metaverse services.

\section{ACKNOWLEDGMENTS}
This work was partially supported by JST ASPIRE Grant Number JPMJAP2327, Japan.


\def\refname{REFERENCES}
\bibliographystyle{IEEEtran}
\bibliography{main_revision}

\begin{thebibliography}{10}
\providecommand{\url}[1]{#1}
\csname url@samestyle\endcsname
\providecommand{\newblock}{\relax}
\providecommand{\bibinfo}[2]{#2}
\providecommand{\BIBentrySTDinterwordspacing}{\spaceskip=0pt\relax}
\providecommand{\BIBentryALTinterwordstretchfactor}{4}
\providecommand{\BIBentryALTinterwordspacing}{\spaceskip=\fontdimen2\font plus
\BIBentryALTinterwordstretchfactor\fontdimen3\font minus \fontdimen4\font\relax}
\providecommand{\BIBforeignlanguage}[2]{{%
\expandafter\ifx\csname l@#1\endcsname\relax
\typeout{** WARNING: IEEEtran.bst: No hyphenation pattern has been}%
\typeout{** loaded for the language `#1'. Using the pattern for}%
\typeout{** the default language instead.}%
\else
\language=\csname l@#1\endcsname
\fi
#2}}
\providecommand{\BIBdecl}{\relax}
\BIBdecl

\bibitem{Slater2022-mr}
M.~Slater, D.~Banakou, A.~Beacco, J.~Gallego, F.~Macia-Varela, and R.~Oliva, ``\BIBforeignlanguage{en}{A separate reality: An update on place illusion and plausibility in virtual reality},'' \emph{\BIBforeignlanguage{en}{Front. Virtual Real.}}, vol.~3, p. 914392, Jun. 2022.

\bibitem{Botin-Sanabria2022-zd}
D.~M. Botín-Sanabria, A.-S. Mihaita, R.~E. Peimbert-García, M.~A. Ramírez-Moreno, R.~A. Ramírez-Mendoza, and J.~d.~J. Lozoya-Santos, ``\BIBforeignlanguage{en}{Digital twin technology challenges and applications: A comprehensive review},'' \emph{\BIBforeignlanguage{en}{Remote Sens. (Basel)}}, vol.~14, no.~6, p. 1335, Mar. 2022.

\bibitem{Ritterbusch2023-ct}
G.~D. Ritterbusch and M.~R. Teichmann, ``\BIBforeignlanguage{en}{Defining the metaverse: A systematic literature review},'' \emph{\BIBforeignlanguage{en}{IEEE Access}}, vol.~11, pp. 12\,368--12\,377, 2023.

\bibitem{Freeman2021-ad}
G.~Freeman and D.~Maloney, ``\BIBforeignlanguage{en}{Body, avatar, and me: The presentation and perception of self in social virtual reality},'' \emph{\BIBforeignlanguage{en}{Proc. ACM Hum. Comput. Interact.}}, vol.~4, no. CSCW3, pp. 1--27, Jan. 2021.

\bibitem{Suh2018-ai}
M.~m. Suh, F.~Bentley, and D.~Lottridge, ``\BIBforeignlanguage{en}{It's kind of boring looking at just the face: How teens multitask during mobile videochat},'' \emph{\BIBforeignlanguage{en}{Proc. ACM Hum. Comput. Interact.}}, vol.~2, no. CSCW, pp. 1--23, Nov. 2018.

\bibitem{Slater2016-tp}
M.~Slater and M.~V. Sanchez-Vives, ``\BIBforeignlanguage{en}{Enhancing our lives with immersive virtual reality},'' \emph{\BIBforeignlanguage{en}{Front. Robot. AI}}, vol.~3, p. 236866, Dec. 2016.

\bibitem{Milgram1995-ku}
P.~Milgram, H.~Takemura, A.~Utsumi, and F.~Kishino, ``Augmented reality: a class of displays on the reality-virtuality continuum,'' in \emph{Telemanipulator and Telepresence Technologies}, H.~Das, Ed.\hskip 1em plus 0.5em minus 0.4em\relax SPIE, Dec. 1995, pp. 282--292.

\bibitem{Zeltzer1992-ku}
D.~Zeltzer, ``Autonomy, interaction, and presence,'' \emph{Presence: Teleoperators and Virtual Environments}, vol.~1, no.~1, pp. 127--132, 02 1992.

\bibitem{Abich2021-ht}
J.~Abich, IV, J.~Parker, J.~S. Murphy, and M.~Eudy, ``\BIBforeignlanguage{en}{A review of the evidence for training effectiveness with virtual reality technology},'' \emph{\BIBforeignlanguage{en}{Virtual Real.}}, vol.~25, no.~4, pp. 919--933, Dec. 2021.

\bibitem{Jensen2018-dv}
L.~Jensen and F.~Konradsen, ``\BIBforeignlanguage{en}{A review of the use of virtual reality head-mounted displays in education and training},'' \emph{\BIBforeignlanguage{en}{Educ. Inf. Technol.}}, vol.~23, no.~4, pp. 1515--1529, Jul. 2018.

\bibitem{Xie2021-cb}
B.~Xie, H.~Liu, R.~Alghofaili, Y.~Zhang, Y.~Jiang, F.~D. Lobo, C.~Li, W.~Li, H.~Huang, M.~Akdere, C.~Mousas, and L.-F. Yu, ``\BIBforeignlanguage{en}{A review on virtual reality skill training applications},'' \emph{\BIBforeignlanguage{en}{Front. Virtual Real.}}, vol.~2, p. 645153, Apr. 2021.

\bibitem{Yung2019-km}
R.~Yung and C.~Khoo-Lattimore, ``\BIBforeignlanguage{en}{New realities: a systematic literature review on virtual reality and augmented reality in tourism research},'' \emph{\BIBforeignlanguage{en}{Curr. Issues Tourism}}, vol.~22, no.~17, pp. 2056--2081, Oct. 2019.

\bibitem{Loomis1999-io}
J.~M. Loomis, J.~J. Blascovich, and A.~C. Beall, ``\BIBforeignlanguage{en}{Immersive virtual environment technology as a basic research tool in psychology},'' \emph{\BIBforeignlanguage{en}{Behav. Res. Methods Instrum. Comput.}}, vol.~31, no.~4, pp. 557--564, Nov. 1999.

\bibitem{Maister2015-rg}
L.~Maister, M.~Slater, M.~V. Sanchez-Vives, and M.~Tsakiris, ``\BIBforeignlanguage{en}{Changing bodies changes minds: owning another body affects social cognition},'' \emph{\BIBforeignlanguage{en}{Trends Cogn. Sci.}}, vol.~19, no.~1, pp. 6--12, Jan. 2015.

\bibitem{Blanke2015-tb}
O.~Blanke, M.~Slater, and A.~Serino, ``\BIBforeignlanguage{en}{Behavioral, neural, and computational principles of bodily self-consciousness},'' \emph{\BIBforeignlanguage{en}{Neuron}}, vol.~88, no.~1, pp. 145--166, Oct. 2015.

\bibitem{Herrera2018-mj}
F.~Herrera, J.~Bailenson, E.~Weisz, E.~Ogle, and J.~Zaki, ``\BIBforeignlanguage{en}{Building long-term empathy: A large-scale comparison of traditional and virtual reality perspective-taking},'' \emph{\BIBforeignlanguage{en}{PLoS One}}, vol.~13, no.~10, p. e0204494, Oct. 2018.

\bibitem{Bertrand2018-kw}
P.~Bertrand, J.~Guegan, L.~Robieux, C.~A. McCall, and F.~Zenasni, ``\BIBforeignlanguage{en}{Learning empathy through virtual reality: Multiple strategies for training empathy-related abilities using body ownership illusions in embodied virtual reality},'' \emph{\BIBforeignlanguage{en}{Front. Robot. AI}}, vol.~5, p.~26, Mar. 2018.

\bibitem{Genay2022-pm}
A.~Genay, A.~Lécuyer, and M.~Hachet, ``Being an avatar “for real”: A survey on virtual embodiment in augmented reality,'' \emph{IEEE Trans. Vis. Comput. Graph.}, vol.~28, no.~12, pp. 5071--5090, Dec. 2022.

\bibitem{nick2007yee}
N.~Yee and J.~Bailenson, ``The proteus effect: The effect of transformed self-representation on behavior,'' \emph{Human Communication Research}, vol.~33, no.~3, pp. 271--290, 2007.

\bibitem{Okamoto2015-vw}
T.~Okamoto, ``\BIBforeignlanguage{en}{Otaku tourism and the anime pilgrimage phenomenon in japan},'' \emph{\BIBforeignlanguage{en}{Jpn. Forum}}, vol.~27, no.~1, pp. 12--36, Jan. 2015.

\bibitem{Sessions2010-qu}
L.~F. Sessions, ``\BIBforeignlanguage{en}{{HOW} {OFFLINE} {GATHERINGS} {AFFECT} {ONLINE} {COMMUNITIES}: When virtual community members ‘meetup’},'' \emph{\BIBforeignlanguage{en}{Inf. Commun. Soc.}}, vol.~13, no.~3, pp. 375--395, Apr. 2010.

\bibitem{Liu2023-px}
J.~Liu, ``\BIBforeignlanguage{en}{Virtual presence, real connections: Exploring the role of parasocial relationships in virtual idol fan community participation},'' \emph{\BIBforeignlanguage{en}{Glob. Media China}}, Dec. 2023.

\bibitem{Song2022-hb}
S.~Song, J.~Baba, J.~Nakanishi, Y.~Yoshikawa, and H.~Ishiguro, ``\BIBforeignlanguage{en}{Costume vs. wizard of oz vs. telepresence: How social presence forms of tele-operated robots influence customer behavior},'' \emph{\BIBforeignlanguage{en}{Human-Robot Interaction}}, pp. 521--529, Mar. 2022.

\bibitem{Lu2021-pr}
Z.~Lu, C.~Shen, J.~Li, H.~Shen, and D.~Wigdor, ``\BIBforeignlanguage{en}{More kawaii than a real-person live streamer: Understanding how the otaku community engages with and perceives virtual {YouTubers}},'' in \emph{\BIBforeignlanguage{en}{Proceedings of the 2021 CHI Conference on Human Factors in Computing Systems}}.\hskip 1em plus 0.5em minus 0.4em\relax New York, NY, USA: ACM, May 2021.

\bibitem{Byun2023-db}
K.~J. Byun and S.~J.~g. Ahn, ``\BIBforeignlanguage{en}{A systematic review of virtual influencers: Similarities and differences between human and virtual influencers in interactive advertising},'' \emph{\BIBforeignlanguage{en}{J. Interact. Advert.}}, vol.~23, no.~4, pp. 293--306, Oct. 2023.

\bibitem{Laszkiewicz2023-ws}
A.~Laszkiewicz and M.~Kalinska-Kula, ``\BIBforeignlanguage{en}{Virtual influencers as an emerging marketing theory: A systematic literature review},'' \emph{\BIBforeignlanguage{en}{Int. J. Consum. Stud.}}, vol.~47, no.~6, pp. 2479--2494, Nov. 2023.

\bibitem{hatada2024robotcafe}
Y.~Hatada, G.~Barbareschi, K.~Takeuchi, H.~Kato, K.~Yoshifuji, K.~Minamizawa, and T.~Narumi, ``People with disabilities redefining identity through robotic and virtual avatars: A case study in avatar robot cafe,'' in \emph{Proceedings of the 2024 CHI Conference on Human Factors in Computing Systems}, ser. CHI '24.\hskip 1em plus 0.5em minus 0.4em\relax New York, NY, USA: Association for Computing Machinery, 2024.

\bibitem{Cho2023-bm}
S.~Cho, J.~Lee, and B.~Suh, ``\BIBforeignlanguage{en}{``{I} want to reveal, but {I} also want to hide'' understanding the conflict of revealing and hiding needs in virtual study rooms},'' \emph{\BIBforeignlanguage{en}{Proc. ACM Hum. Comput. Interact.}}, vol.~7, no. CSCW2, pp. 1--27, Sep. 2023.

\bibitem{Lee2021-ap}
Y.~Lee, J.~J.~Y. Chung, J.~Y. Song, M.~Chang, and J.~Kim, ``\BIBforeignlanguage{en}{Personalizing ambience and illusionary presence: How people use “study with me” videos to create effective studying environments},'' in \emph{\BIBforeignlanguage{en}{Proceedings of the 2021 CHI Conference on Human Factors in Computing Systems}}.\hskip 1em plus 0.5em minus 0.4em\relax New York, NY, USA: ACM, May 2021.

\bibitem{Maloney2020-es}
D.~Maloney and G.~Freeman, ``\BIBforeignlanguage{en}{Falling asleep together: What makes activities in social virtual reality meaningful to users},'' in \emph{\BIBforeignlanguage{en}{Proceedings of the Annual Symposium on Computer-Human Interaction in Play}}.\hskip 1em plus 0.5em minus 0.4em\relax New York, NY, USA: ACM, Nov. 2020.

\bibitem{Miura2022-yx}
R.~Miura, S.~Kasahara, M.~Kitazaki, A.~Verhulst, M.~Inami, and M.~Sugimoto, ``\BIBforeignlanguage{en}{{MultiSoma}: Motor and gaze analysis on distributed embodiment with synchronized behavior and perception},'' \emph{\BIBforeignlanguage{en}{Front. Comput. Sci.}}, vol.~4, p. 788014, May 2022.

\end{thebibliography}

\begin{IEEEbiography}{Yuichi Hiroi}{\,} is a senior research scientist at Cluster Metaverse Lab, Japan. His current research interests include augmented reality, near-eye displays and personalized vision augmentation. He received his Ph.D. degree in engineering from Tokyo Institute of Technology. He is a Member of the IEEE Computer Society. Contact him at y.hiroi@cluster.mu.
\end{IEEEbiography}

\begin{IEEEbiography}{Yuji Hatada}{\,}
 is an Assistant Professor at the Interfaculty Initiative in Information Studies, The University of Tokyo. His current research interests include  virtual reality, cognitive science, and narrative transformations that users experience through avatars. He received his Ph.D. degree in interdisciplinary information studies in 2023 from the University of Tokyo. Contact him at hatada@nae-lab.org.
\end{IEEEbiography}

\begin{IEEEbiography}{Takefumi Hiraki} {\,} is a senior research scientist at Cluster Metaverse Lab and an associate professor at the University of Tsukuba, Japan. His current research interests include augmented reality, haptic interfaces, and soft robotics. He received his Ph.D. degree in engineering from the University of Tokyo. He is a Member of the IEEE Computer Society. Contact him at t.hiraki@cluster.mu.
\end{IEEEbiography}

\end{document}